\documentclass[12pt,preprint]{aastex}

\shorttitle{Crystalline Silicate around the HD145263}
\shortauthors{Honda et al.}

\begin{document}

\title{Crystalline Silicate Feature of the Vega-like star HD145263 \altaffilmark{1}}


\author{Mitsuhiko Honda\altaffilmark{2,3,4}, 
Hirokazu Kataza\altaffilmark{4}, 
Yoshiko K. Okamoto\altaffilmark{5}, Takashi Miyata\altaffilmark{6},
Takuya Yamashita\altaffilmark{3,2}, 
Shigeyuki Sako\altaffilmark{7,3}, Takuya Fujiyoshi\altaffilmark{3},
Meguru Ito\altaffilmark{2}, Yoko Okada\altaffilmark{2}, Itsuki
Sakon\altaffilmark{2} and Takashi Onaka\altaffilmark{2}}


\altaffiltext{1}{Based on data collected at Subaru Telescope, which is operated
by the National Astronomical Observatory of Japan.}
\altaffiltext{2}{Department of Astronomy, School of Science, University
of Tokyo, Bunkyo-ku, Tokyo 113-0033, Japan, hondamt@ir.isas.jaxa.jp,
onaka@astron.s.u-tokyo.ac.jp}
\altaffiltext{3}{Subaru Telescope, National Astronomical Observatory of
Japan, 650 North A'ohoku Place, Hilo, Hawaii 96720, U.S.A.,
takuya@naoj.org}
\altaffiltext{4}{Department of Infrared Astrophysics, Institute of Space
and Astronautical Science, Japan Aerospace Exploration Agency, Yoshinodai
3-1-1, Sagamihara, Kanagawa, 229-8510, Japan, kataza@ir.isas.jaxa.jp}
\altaffiltext{5}{Center for Natural Sciences, College of Liberal Arts and
Sciences, Kitasato University, 1-15-1 Kitazato, Sagamihara, Kanagawa
228-8555, Japan, okamtoys@cc.nao.ac.jp}
\altaffiltext{6}{Kiso Observatory, Institute of Astronomy, School of
Science, University of Tokyo, Mitake, Nagano 397-0101, Japan, 
miyata@kiso.ioa.s.u-tokyo.ac.jp}
\altaffiltext{7}{Institute of Astronomy, University of Tokyo, 2-21-1
Osawa, Mitaka, Tokyo 181-0015, Japan, sako@ioa.s.u-tokyo.ac.jp}


\begin{abstract}
We have observed the 8-13\,$\mu$m spectrum (R$\sim$250) of the Vega-like
 star candidate HD145263 using Subaru/COMICS. 
The spectrum of HD145263 shows the broad trapezoidal silicate
 feature with the shoulders at 9.3\,$\mu$m and 11.44\,$\mu$m,
 indicating the presence of crystalline silicate grains. This detection
 implies that crystalline silicate may also be commonly present
 around Vega-like stars.
 The 11.44\,$\mu$m feature is slightly shifted to a longer
 wavelength compared to the usual 11.2-3\,$\mu$m crystalline forsterite
 feature detected toward Herbig Ae/Be stars and T Tauri stars. 
 Although the peak shift due to the effects of the grain size
 can not be ruled out, we suggest that Fe-bearing crystalline
 olivine explains the observed peak wavelength fairly well.
 Fe-bearing silicates are commonly found in
 meteorites and most interplanetary dust particles, which originate from
 planetesimal-like asteroids. According to studies of
 meteorites, Fe-bearing silicate must have been formed in asteroidal
 planetesimals, supporting the scenario that dust grains around
 Vega-like stars are of planetesimal origin, if the observed
 11.44\,$\mu$m peak is due to Fe-bearing silicates.
\end{abstract}


\section{Introduction}
Recent mid-infrared observations have revealed that crystalline silicates are
present in circumstellar disks around young stars
\citep[e.g.][]{hanner95, malfait98, sitko99, meeus01, bouwman01,
vanboekel03, honda03, meeus03, przygodda03}. Observed dust
features indicate that the major composition of crystalline
silicate is Mg-pure olivine, forsterite (Mg$_2$SiO$_4$) 
\citep[e.g.][]{malfait98}. The observed 69\,$\mu$m
feature strongly supports the presence of forsterite
\citep{molster02,bowey02} and places a strong upper limit on the Fe-content
in crystalline silicate grains (Fe/(Mg+Fe)$<$5\%). 
Furthermore, the remarkable similarity between the spectrum of the isolated
Herbig Ae/Be star HD100546 and that of the comet Hale-Bopp indicates
that cometary dust is also Mg-rich silicate
\citep{crovisier97,malfait98,wooden99}. 
This result is consistent with the {\it in situ} measurements of Comet
Halley dust by PUMA mass spectrometry \citep{brownlee87}. Further,
\cite{bradley99} reported that interplanetary dust
particles (IDPs) originated from comets, 
such as the anhydrous chondritic porous (CP)
``pyroxene'' class of IDPs, contain mostly Mg-pure silicate grains
(forsterite and enstatite). 
On the other hand, chondritic meteorites and most IDPs which have 
records of early solar system processing,
show a variety of silicates. Not only Mg-rich olivine but also
Fe-bearing fayalitic olivine are commonly present. 
In meteorites which probably came
from asteroids, Fe-bearing silicate is
ubiquitos \citep[e.g.][]{krot00}. Many IDPs, which originate from either
asteroids or comets, also show various Mg-Fe silicates \citep{bradley99}.
However, there is no evidence for Fe-bearing crystalline silicate grains
from astronomical observations so far \citep{suto02}. 

Debris dust disk around Vega-like stars are supposed to be
continuously replenished from planetesimals and/or comets
\citep{lagrange00}. Thus one can expect that not only Mg-rich crystalline
silicate but also Fe-bearing crystalline silicate grains might be
present in the debris disk.
Among Vega-like stars, compositional studies of silicate dust 
are rather scarce. The presence of crystalline silicate is reported only for
$\beta$Pic \citep{knacke93,weinberger03,okamoto04}, and no Fe-bearing
crystalline silicate grains are reported.

In this Letter, we show the 8-13\,$\mu$m spectrum of the Vega-like
star candidate HD145263, and present a possible hint for Fe-bearing
crystalline silicate. HD145263 is an F0V star at a distance of 116 pc
\citep{sylvester00}, and was originally reported as a Vega-like star
candidate by \cite{mannings98}. It is located close to the zero-age
main-sequence (ZAMS) in the HR diagram \citep{sylvester00}. It has
L$_{IR}$/L$_{*}$ of 0.02 which is smaller than typical Herbig Ae/Be
stars and T Tauri stars, but larger than prototype Vega-like stars 
\citep{lagrange00}. 
Thus it could be a young Vega-like star. It is also a member of the Upper
Scorpius association whose age is estimated to be 8-10 Myr
\citep{sartori03}, further supporting that HD145263 is a young
Vega-like star.

\section{Observations and Data Reduction}
HD145263 was observed with the Cooled Mid-Infrared Camera and Spectrometer 
\citep[ COMICS; ][]{kataza00, okamoto03, sako03} mounted on the 8.2m SUBARU
Telescope on July 15, 2003. N-band low-resolution (R$\sim$250) spectroscopic
observations were performed with the 0.33'' wide slit.
Imaging photometry observations in the 8.8\,$\mu$m ($\Delta\lambda$=0.8\,$\mu$m) and 12.4\,$\mu$m ($\Delta\lambda$=1.2\,$\mu$m) bands were also made.
We selected HD133774 \citep{cohen99} as the ratioing star for the
correction of the atmospheric absorption and as the flux standard star for
the aperture photometry. The absolute flux of the standard
star is obtained by integrating the template spectra provided by \cite{cohen99}. 
The observation parameters are summarized in Table \ref{tbl-1}.

Standard reduction procedures as described in \cite{honda03,honda04}
were applied for aperture photometry and spectroscopic data. 
The wavelength uncertainty is estimated to be 0.0025\,$\mu$m
\citep{okamoto03}. The effect of different airmass between the object and
the standard star was corrected using the atmospheric transmission spectra
calculated by the ATRAN software \citep{lord92}.
Finally we adjusted the flux of the spectrum of the
HD145263 to the flux measured in the 8.8\,$\mu$m and 12.4\,$\mu$m bands
respectively to correct the slit throughput. 
Detailed descriptions of the reduction procedures are given in
\cite{honda03,honda04}

\section{Results}
\begin{figure}
\plotone{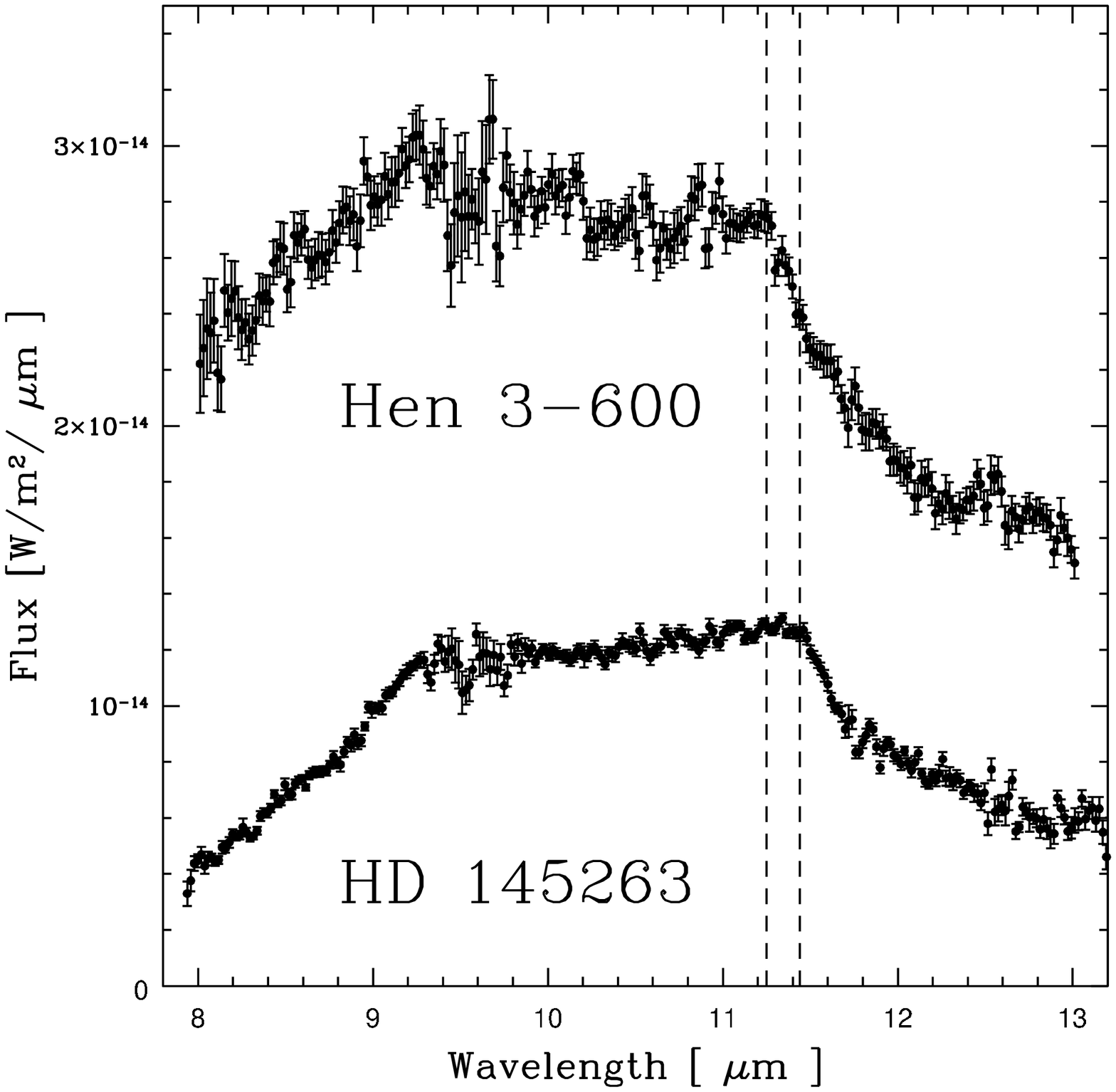}
\caption{Observed 8-13\,$\mu$m spectrum of the Vega-like star candidate
 HD145263 (lower spectrum) together with the T Tauri star Hen 3-600A
 (upper spectrum) for comparison \citep{honda03}. \label{fig1}
Large scatter in the 9.3-9.9\,$\mu$m region is due to the 
atmospheric ozone absorption.
HD145263 shows the 11.44\,$\mu$m shoulder, while Hen 3-600A shows
 the 11.24\,$\mu$m forsterite feature. }
\end{figure}

\begin{figure}
\plotone{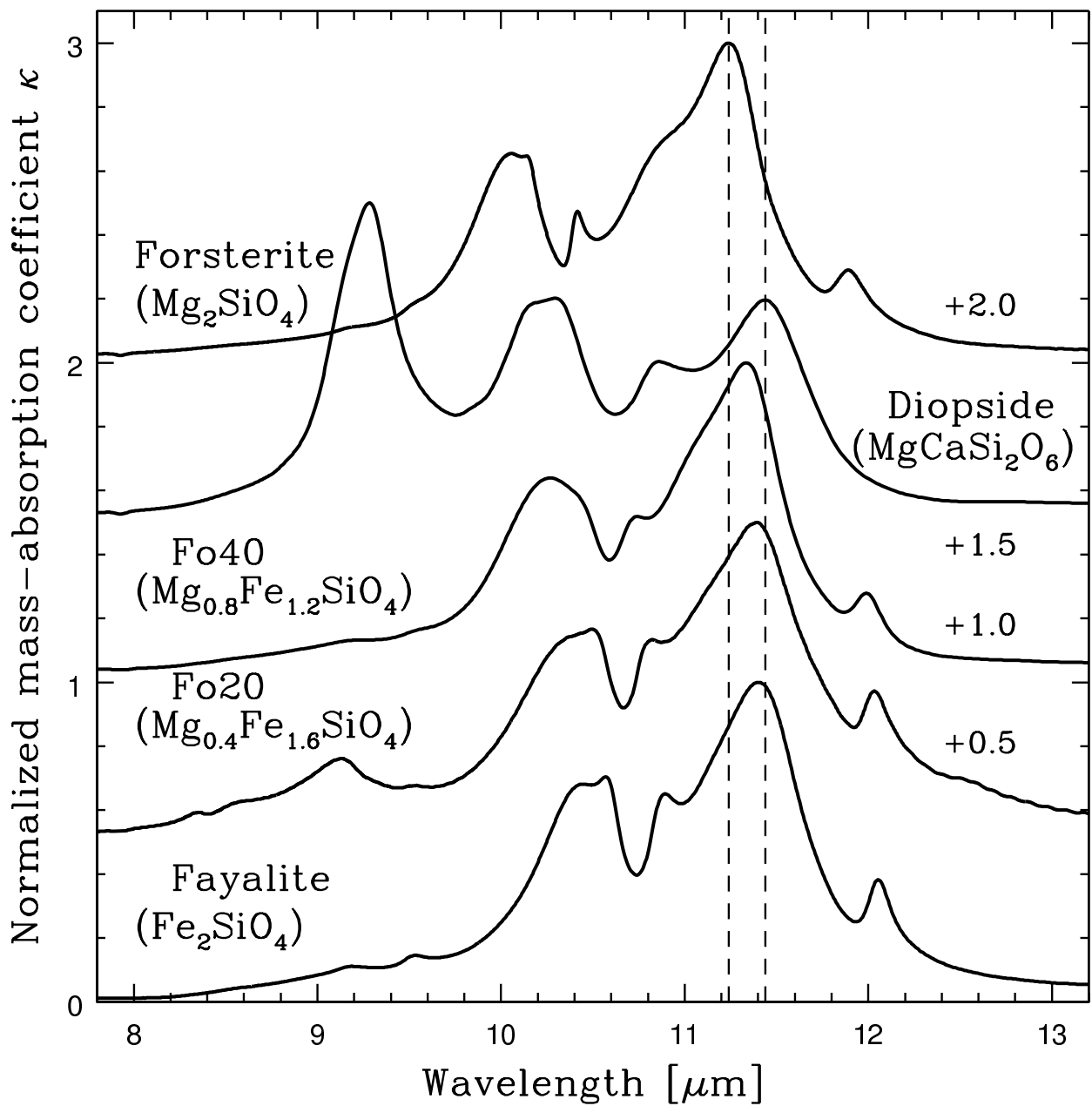}
\caption{Normalized spectral profiles of the candidate dust species 
for the spectrum of HD145263\label{fig2}. The data of
 forsterite, Fo40, Fo21.8, fayalite \citep{koike03} and diopside
 \citep{koike00} are taken from laboratory absorption measurements. }
\end{figure}

Figure \ref{fig1} shows the observed 8-13\,$\mu$m spectra of HD145263
together with the T Tauri star Hen 3-600A \citep{honda03} for comparison.
HD145263 shows a broad trapezoidal silicate emission feature with the
shoulders at 9.3\,$\mu$m and 11.44\,$\mu$m. This shape indicates
the presence of crystalline silicate dust species in this system. Thus this
is the second detection of crystalline silicate grains around Vega-like
stars after $\beta$ Pic.
However, the spectrum is rather smooth between 9.3\,$\mu$m and
11.44\,$\mu$m, which is unusual for typical crystalline silicate
features. According to laboratory measurements of fine grained
crystalline forsterite particles \citep{koike03}, the sub-peak at
10.1\,$\mu$m should be observed in addition to the 11.24\,$\mu$m strong
peak if it is attributed to forsterite grains (see Fig.\ref{fig2}). 
Furthermore, the 11.44\,$\mu$m peak is
clearly shifted to a longer wavelength compared to the forsterite
11.24\,$\mu$m feature that has been seen toward many Herbig Ae/Be stars
and T Tauri stars.
The 11.44\,$\mu$m feature may be attributed to
1) the presence of other dust species which account for the
feature, or
2) the peak shift of the 11.24\,$\mu$m forsterite feature due to the effect
 of the dust temperature, shape, and/or size.

\begin{figure}
\plotone{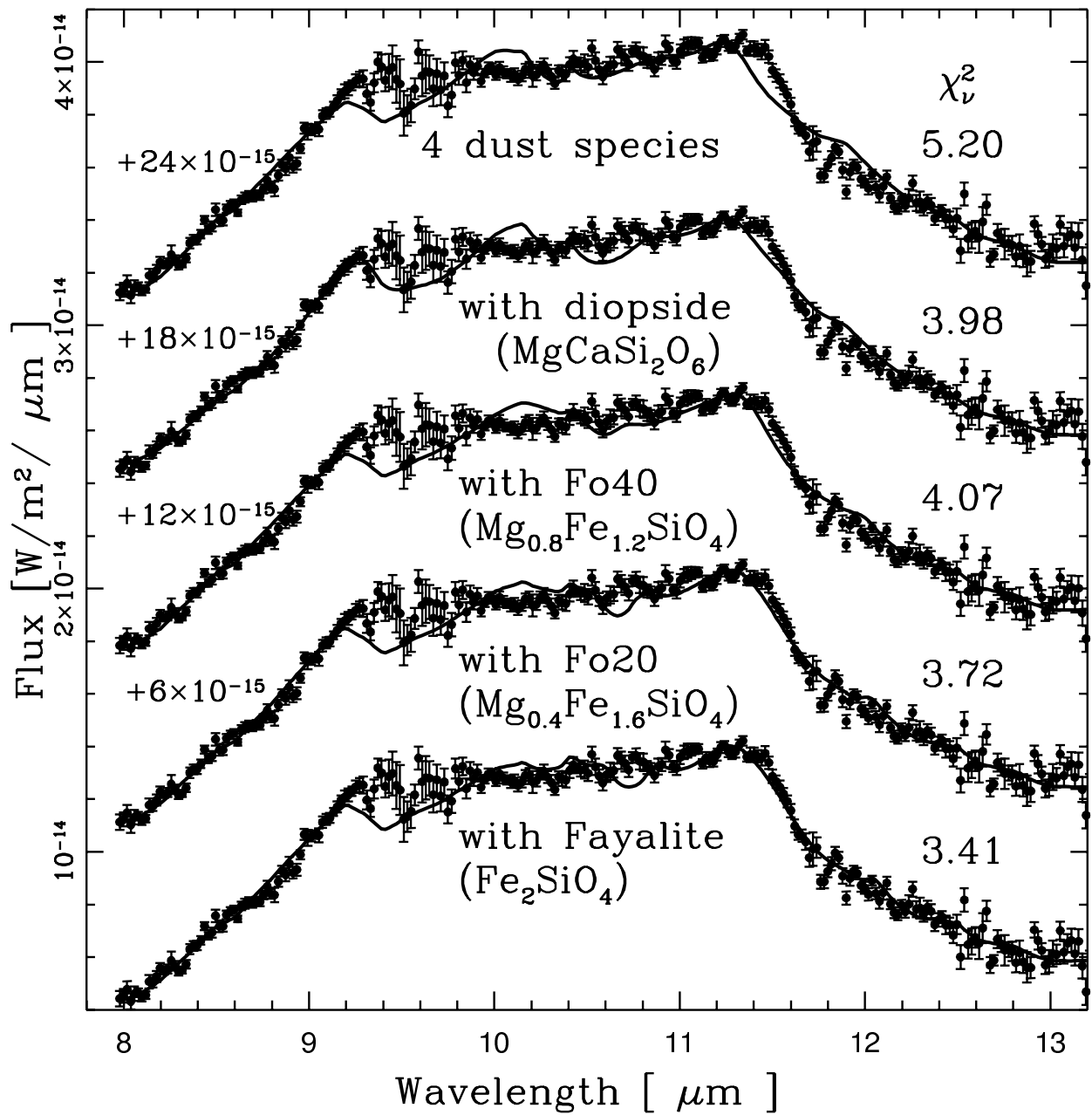}
\caption{Observed spectrum of HD145263 together with model fit results of
 various dust combinations. The solid lines indicate best fit result
 for the each dust combination as described in Table \ref{tbl-2}. 
 As seen in this figure, the dust combination included with fayalitic
 olivine provide the best fit result. \label{fig3}}
\end{figure}

Candidate dust species which might account for the 11.44\,$\mu$m feature
 are, for example, fayalitic olivine (e.g. Fo20 or fayalite in
 \cite{koike03}) and diopside \citep{koike00}. They have features at
 around 11.4\,$\mu$m. In order to reproduce the observed
11.44\,$\mu$m feature and overall silicate emission of HD145263,
we made a simple profile fitting. The fitting formula is given by
$$\lambda F_\lambda=(a_0+\sum_{i=1}^{} a_i\kappa_i)
 (\frac{\lambda}{9.8[\mu m]})^n,$$
where $\kappa_i$ is mass absorption coefficient of dust and $a_i$ are
 the fitting parameters.
At first, we considered the following 4 dust species that were
originally proposed by \cite{bouwman01} for reproducing the silicate
 feature of Hebig Ae/Be stars : 0.1\,$\mu$m and 2.0\,$\mu$m glassy olivine
\citep[MgFeSiO$_4$; Mie calculations using optical constant by][]{dorschner95}, crystalline forsterite
\citep[Mg$_2$SiO$_4$; absorption measurements by][]{koike03}, and silica
 \citep[SiO$_2$; CDE calculations using optical constant by][]{spitzer61} grains. Their spectral profiles 
are shown in \cite{honda03}.
Resulting best-fit model spectra are shown in Figure \ref{fig3} and
the reduced $\chi$ squared ($\chi_\nu^2$)
are summarized in Table \ref{tbl-2}. 
The combination of the ``standard'' 4 dust species can
not provide acceptable fitting results. The 11.44\,$\mu$m shoulder is
 not well reproduced.
To improve the model spectrum, we introduce the fifth component : 
diopside \citep{koike00}, Fo40,
Fo21.8, and fayalite \citep{koike03} (Figure \ref{fig2}). The fitting
results are also summarized in Table \ref{tbl-2} and Figure \ref{fig3}.
From Table \ref{tbl-2}, fayalitic olivine provides the best fit
result. This is because fayalitic olivine has a feature at 11.4\,$\mu$m
and makes the total spectrum between 9.3 and 11.44\,$\mu$m
flatter if it coexists with forsterite.
The presence of fayalitic olivine can account for the feature in the
10\,$\mu$m spectrum of HD145263 fairly well.

On the other hand, there are alternative possibilities that could account
for the 11.44\,$\mu$m feature. 
The peak wavelengths of crystalline silicate features
depend on the dust physical parameters \citep{molster02} such as the
temperature \citep{chihara01,bowey01}, the shape, and the size
\citep[e.g.][]{fabian01,min03}. 
The feature wavelength generally shifts to longer wavelengths as the
grain temperature increases. However, \cite{henning97} reported that the
temperature effect for the crystalline silicate (bronzite) feature at
10\,$\mu$m band is very small.
Recent preliminary laboratory measurements show that the amount of peak shift
($\Delta\lambda$) between 50K and 300K for the 11.24\,$\mu$m
forsterite feature is about $\sim$0.01\,$\mu$m (Suto et al., private
communication), indicating that the temperature effect only can not
explain the 11.44\,$\mu$m feature even at the forsterite sublimation
temperature ($\sim$1400\,K, $\Delta\lambda < 0.2\,\mu$m ).
Further, lack of excess emission in the near
infrared from the spectral energy distribution of the HD145263
\citep{sylvester00} indicates that there is little hot dust and 
it is difficult to expect the peak shift due to the grain temperature effect.
Very elongated forsterite ellipsoids (prolates of aspect ratio of 10
$\sim$ 100) have the feature around 11.4\,$\mu$m \citep{fabian01}, 
but the presence of such grain shape is unlikely according to the
experimental study of fragments from collisional disruption
\citep{fujiwara78,capaccioni84,fujiwara89}. Large ( a few
micron) crystalline forsterite grains have the possibility to account for the
11.44\,$\mu$m shoulder. Large grains tend to broaden the feature and
shift the peak to a longer wavelength \citep{molster02}. 
Although detailed studies of optical properties of large ellipsoidal
crystalline silicate grains are needed to quantitatively investigate
this possibility, large crystalline
forsterite grains seem to be able to account for the peak shift to the
$\sim$11.4\,$\mu$m \citep{min04}.

To summarize, the observed 11.44\,$\mu$m feature detected in HD145263
can be accounted for by either size or Fe-inclusion effects of olivine
grains, or both. While the presence of large grains is not unexpected for
Vega-like stars since its dust is replenished from much larger bodies,
the presence of Fe-bearing silicate is also likely to occur.
In the following section, we focus on the possibility and consequence
of Fe-bearing olivine grains.

\section{Discussion}

Silicate material in planetesimals should undergo substantial alterations
by thermal metamorphism through dust accumulating process or radioactive
decay of species like $^{26}$Al in them \citep[e.g.][]{huss01}. Such
alterations are likely to lead to (partial) crystallization of silicate.
Thus one can expect that crystalline silicate dust may appear when
grains are supplied from the collisions of planetesimals, as is the case
for the Vega-like stars. A similar origin for the crystalline silicate
dust around the old Herbig Ae/Be star HD100546 and the comet Hale Bopp was
proposed by \cite{bouwman03}. 
The presence of crystalline silicate dust around the Vega-like star
HD145263 supports the above hypothesis
that crystalline silicate grains may be common in debris disks.
However, this situation seems not to be the case for $\beta$Pic, which is
another Vega-like star with a signature of crystalline silicate
\citep{okamoto04}. The crystalline silicate grains around $\beta$Pic
are suggested to be formed in the vicinity of the central star, but not in the
planetesimals. Thus the origin for the crystalline silicate grains
around Vega-like stars seem to have a variety.

HD145263 spectrum indicates a possible 
presence of Fe-bearing crystalline silicates, while the crystalline
 silicate in the younger precursors such as T Tauri stars and Herbig
 Ae/Be stars is Mg-pure or Mg-rich, supporting the asteroidal
 plantesimal origin for crystalline silicate grains around HD145263.
What is the origin of the Fe-bearing crystalline silicates?
The dust around Vega-like stars is supposed to be of asteroidal planetesimals
and comets \citep{lagrange00}. Cometary silicate grains are indicated
to be Mg-rich based on the {\it in situ} measurements of the
comet Halley dust \citep{brownlee87}, infrared observations of
comet Hale Bopp \citep{crovisier97,malfait98,wooden99}, and studies of
the anhydrous chondritic porous ``pyroxene'' class of IDPs which may be
of cometary origin \citep{bradley99}. On the other hand,
among the meteorites which came from asteroids,
the presence of Fe-bearing silicate material is quite common
\citep[e.g.][]{krot00} and Mg-pure silicate (forsterite and enstatite)
grains are minor constituents of them \citep{bradley99}. 
According to studies of meteorites \citep[][and references therein]{krot00}, 
fayalitic olivine in meteorites is likely to have been formed in 
planetesimal-like parent bodies by low-temperature ($\sim$500K)
alteration under the presence of aqueous solution, because the formation of
fayalitic olivine in a gas of solar composition is kinetically
difficult or inhibited.
Though condensation from a gas more oxidized than the solar gas may also lead
to fayalitic olivine formation, such processes have difficulties to
account for all the observed characteristics of meteorites \citep{krot00}.
If fayalitic olivine is really formed in the asteroidal planetesimals, 
the presence of fayalitic olivine grains suggets mineralogical
evidence that grains are indeed of asteroidal planetesimal origin. 

To make solid detection of Fe-bearing olivine grains, 
observations at longer wavelengths are crucial.
Accroding to \cite{koike03}, similar peak shifts to longer wavelengths 
for longer wavelength bands ($>20$\,$\mu$m) should be observed when Fe-bearing crystalline silicate grains
really exist. Observations using Spitzer Space Telescope and Astro-F
are strongly desired to provide decisive data.

\acknowledgments
We are grateful to all of the staff members of SUBARU Telescope for
providing us the opportunities for these observations and their supports. 
We thank Chiyoe Koike and Hiroki Chihara for providing us
crystalline silicates spectra and useful comments. We also thank Hiroko
Nagahara, Hiroshi Suto, and Sunao Hasegawa for fruitful discussion.
M. H. is financially supported by the Japan Society for the Promotion of
Science.

\clearpage

\begin{deluxetable}{cccccc}
\tabletypesize{\scriptsize}
\tablecaption{COMICS Observations of the Vega-like star candidate HD145263\label{tbl-1}}
\tablewidth{0pt}
\tablehead{
\colhead{Date} & \colhead{Mode} & \colhead{Object}   & \colhead{Filter [\,$\mu$m]}   &
\colhead{Integ. Time[sec]} & \colhead{Air Mass} 
}

\startdata
July 15, 2003 &  Imaging     & HD145263 &8.8($\Delta\lambda =$0.8) & 40.2 & 1.427\\
          &              & HD145263 &12.4($\Delta\lambda =$1.2)& 60.9 & 1.425 \\
          &              & HD133774 &8.8($\Delta\lambda =$0.8) & 5.0 & 1.452\\
          &              & HD133774 &12.4($\Delta\lambda =$1.2)& 4.1 & 1.462\\
          & Spectroscopy & HD145263 & -                        &1898.4  &1.423-1.444\\
          &              & HD133774 & -                        & 10.8 & 1.440 \\
\enddata
\end{deluxetable}

\begin{deluxetable}{ccccccccc}
\tabletypesize{\scriptsize}
\tablecaption{Combination of dust species and reduced $\chi$ squared
 ($\chi^2_\nu$) for best fit\label{tbl-2}}
\tablewidth{0pt}
\tablehead{
\colhead{0.1\,$\mu$m a. oliv.} & \colhead{2.0\,$\mu$m a. oliv.} &
 \colhead{forsterite}   & \colhead{silica}
 & \colhead{diopside} & \colhead{Fo40}
 & \colhead{Fo21.8} & \colhead{fayalite}& \colhead{$\chi_\nu^2$}
}

\startdata

$\surd$ & $\surd$ & $\surd$ & $\surd$ & & & & & 5.20 \\
$\surd$ & $\surd$ & $\surd$ & $\surd$ & $\surd$ & & & & 3.98 \\
$\surd$ & $\surd$ & $\surd$ & $\surd$ & & $\surd$ & & & 4.07 \\
$\surd$ & $\surd$ & $\surd$ & $\surd$ & & & $\surd$ & & 3.72 \\
$\surd$ & $\surd$ & $\surd$ & $\surd$ & & & & $\surd$ & 3.41 \\
\enddata

\end{deluxetable}


\begin{thebibliography}{}
\bibitem[Bouwman et al.(2001)]{bouwman01} Bouwman, J., Meeus, G., de Koter, A., Hony, S., Dominik, C., \& Waters, L. B. F. M. 2001, \aap, 375, 950
\bibitem[Bouwman et al.(2003)]{bouwman03} Bouwman, J., de Koter, A., Dominik, C., \& Waters, L. B. F. M. 2003, \aap, 401, 577
\bibitem[Bowey et al.(2002)]{bowey02} Bowey, J. E., Barlow, M. J., Molster, F. J., Hofmeister, A. M., Lee, C., Tucker, C., Lim, T., Ade, P. A. R., \& Waters, L. B. F. M., 2002, \mnras, 331, L1
\bibitem[Bowey et al.(2001)]{bowey01} Bowey, J. E., Lee, C., Tucker, C., Hofmeister, A. M., Ade, P. A. R., Barlow, M. J., 2001, \mnras, 325, 886
\bibitem[Bradley et al.(1999)]{bradley99} Bradley, J. P., Snow, T. P., Brownlee, D. E., \& Hanner, M. S. 1999, in Solid Interstellar Matter: The ISO Revolution, ed. by L. d'Hendecourt, C. Joblin, \& A. Jones (EDP Sciences and Springer-Verlag), p297
\bibitem[Brownlee et al.(1987)]{brownlee87} Brownlee, D. E., Wheelock, M. M., Temple, S., Bradley, J. P., \& Kissel, J. 1987, Lunar Planet. Sci., 18, 133
\bibitem[Capaccioni et al.(1984)]{capaccioni84} Capaccioni, F., Cerroni, P., Coradini, M., Farinella, P., Flamini, E., Martelli, G., Paolicchi, P., Smith, P. N., \& Zappala, V. 1984, \nat, 308, 832
\bibitem[Chihara et al.(2001)]{chihara01} Chihara, H., Koike, C., \& Tsuchiyama, A. 2001, \pasj, 53, 243
\bibitem[Cohen et al.(1999)]{cohen99} Cohen, M., Walker, R. G., Carter, B., Hammersley, P., Kidger, M., \& Noguchi, K. 1999, \apj, 117, 1864
\bibitem[Crovisier et al.(1997)]{crovisier97} Crovisier, J., Leech, K., Bockelee-Morvan, D., Brooke, T. Y., Hanner, M. S., Altieri, B., Keller, H. U., \& Lellouch, E. 1997, Science, 275, 1904
\bibitem[Dorschner et al.(1995)]{dorschner95} Dorschner, J., Begemann, B., Henning, Th., J\"{a}ger, C., \& Mutschke, H. 1995, \aap, 300, 503
\bibitem[Fabian et al.(2001)]{fabian01} Fabian, D., Henning, T., J\"{a}ger, C., Mutschke, H., Dorschner, J., \& Wehrhan, O. 2001, \aap, 378, 228
\bibitem[Fujiwara et al.(1978)]{fujiwara78} Fujiwara, A., Kamimoto, G., \& Tsukamoto, A. 1978, \nat, 272, 602
\bibitem[Fujiwara et al.(1989)]{fujiwara89} Fujiwara, A., Cerroni, P., Davis, D., Ryan, E., \& di Martino, M. 1989, in Asteroids II, ed. T. Gehrels (Tuscon: Univ. Arizona Press), 240
\bibitem[Hanner et al.(1995)]{hanner95} Hanner, M. S., Brooke, T. Y., \& Tokunaga, A. T. 1995, \apj, 438, 250
\bibitem[Henning \& Mutschke (1997)]{henning97} Henning, T., \& Mutschke, H. 1997, \aap, 327, 743
\bibitem[Honda et al.(2003)]{honda03} Honda, M., Kataza, H., Okamoto, Y. K., Miyata, T., Yamashita, T., Sako, S., Takubo, S., \& Onaka, T. 2003, \apj, 585, L59
\bibitem[Honda et al.(2004)]{honda04} Honda, M., Watanabe, J., Yamashita, T., Kataza, H., Okamoto, Y., Miyata, T., Sako, S., Fujiyoshi, T., Kawakita, H., Furusho, R., Kinoshita, D., Sekiguchi, T., Ootsubo, T., \& Onaka, T. 2004, \apj, 601, 577
\bibitem[Huss al.(2001)]{huss01} Huss, G. R., MacPherson, G. J., Wasserburg, G. J., Russell, S. S., Srinivasan, G. 2001, Meteoritics and Planetary Science, 36, 975
\bibitem[Kataza et al.(2000)]{kataza00}
Kataza, H., Okamoto, Y., Takubo, S., Onaka, T., Sako, S., Nakamura, K., Miyata, T., \& Yamashita, T. 2000, Proc. SPIE, 4008, 1144
\bibitem[Knacke et al.(1993)]{knacke93}  Knacke, R. F., Fajardo-Acosta, S. B., Telesco, C. M., Hackwell, J. A., Lynch, D. K., \& Russell, R. W. 1993, \apj, 418, 440
\bibitem[Koike et al.(2000)]{koike00} Koike, C., Tsuchiyama, A.,Shibai, H., Suto, H., Tanabe, T., Chihara, H., Sogawa, H., Mouri, H., \& Okada, K., 2000, \aap, 363, 1115
\bibitem[Koike et al.(2003)]{koike03} Koike, C., Chihara, H., Tsuchiyama, A., Suto, H., Sogawa, H., \& Okuda, H. 2003, \aap, 399, 1101
\bibitem[Krot et al.(2000)]{krot00} Krot, A. N., Fegley, B., Jr., Lodders, K., \& Palme, H. 2000, in Protostars \& Planets IV, ed. Mannings, V., Boss, A. P., \& Russell, S. S. (Tuscon: Univ. Arizona Press), p1019
\bibitem[Lagrange et al.(2000)]{lagrange00} Lagrange, A. M., Backman, D. E., \& Artymowicz, P. 2000, in Protostars \& Planets IV, ed. Mannings, V., Boss, A. P., \& Russell, S. S. (Tuscon: Univ. Arizona Press), p639
\bibitem[Lord (1992)]{lord92} Lord, S. D., A New Software Tool for Computing Earth's Atmospheric Transmission of Near- and Far-Infrared Radiation, NASA Technical Memorandum. 103957, Ames Research Center, Moffett Field, California.
\bibitem[Malfait et al.(1998)]{malfait98} Malfait, K., Waelkens, C., Waters, L. B. F. M., Vandenbussche, B., Huygen, E., \& de Graauw, M. S. 1998, \aap, 332, L25
\bibitem[Mannings and Barlow(1998)]{mannings98} Mannings, V., \& Barlow, M. J. 1998, \apj, 497, 330
\bibitem[Meeus et al.(2001)]{meeus01} Meeus, G., Waters, L. B. F. M., Bouwman, J., van den Ancker, M. E., Waelkens, C., \& Malfait, K. 2001, \aap, 365, 476
\bibitem[Meeus et al.(2003)]{meeus03} Meeus, G., Sterzik, M., Bouwman, J., \& Natta, A. 2003, \aap, 409, L25
\bibitem[Min et al.(2003)]{min03} Min, M., Hovenier, J. W., \& de Koter, A. 2003, \aap, 404, 35
\bibitem[Min et al.(2004)]{min04} Min, M., et al. 2004, in preparation
\bibitem[Molster et al.(2002)]{molster02} Molster, F. J., Waters, L. B. F. M., Tielens, A. G. G. M., Koike, C., \& Chihara, H. 2002, \aap, 382, 241
\bibitem[Okamoto et al.(2003)]{okamoto03} Okamoto, Y. K., Kataza, H., Yamashita, T., Miyata, T., Sako, S., Takubo, S., Honda, M., \& Onaka, T. 2003, Proc. of SPIE, 4841, pp 169-180
\bibitem[Okamoto et al.(2004)]{okamoto04} Okamoto, Y. K., Kataza, H., Honda, M., Yamashita, T., Onaka, T., Watanabe, J., Miyata, T., Sako, S., Fujiyoshi, T., \& Sakon, I. 2004, submitted to Nature
\bibitem[Przygodda et al.(2003)]{przygodda03} Przygodda, F., van Boekel, R., Abraham, P., Melnikov, S. Y., Waters, L. B. F. M., \& Leinert, Ch. 2003, \aap, 412, L43
\bibitem[Sako et al.(2003)]{sako03} Sako, S., Okamoto, Y. K., Kataza, H., Miyata, T., Takubo, S., Honda, M., Fujiyoshi, T., Onaka, T., \& Yamashita, T., 2003, \pasp, 115, 1407
\bibitem[Sartori et al.(2003)]{sartori03} Sartori, M. J., Lepine, J. R. D., \& Dias, W. S. 2003, \aap, 404, 913
\bibitem[Sitko et al.(1999)]{sitko99} Sitko, M. L., Grady, C. A., Lynch, D. K., Russel, R. W., \& Hanner, M. S., 1999, \apj, 510, 408
\bibitem[Spitzer \& Kleinman(1961)]{spitzer61} Spitzer, W. G. \& Kleinman, D. A. 1961, Phys. Rev. , 121, 1324
\bibitem[Suto et al.(2002)]{suto02} Suto, H., Koike, C., Sogawa, H., Tsuchiyama, A., Chihara, H., \& Mizutani, K. 2002, \aap, 389, 568
\bibitem[Sylvester et al.(2000)]{sylvester00} Sylvester, R. J., Mannings, V. 2000 \mnras, 313, 73
\bibitem[Sylvester et al.(2001)]{sylvester01} Sylvester, R. J., Dunkin, S. K., \& Barlow, M. J. 2001 \mnras, 327, 133
\bibitem[van Boekel et al.(2003)]{vanboekel03} van Boekel, R., Waters, L. B. F. M., Dominik, C., Bouwman, J., de Koter, A., Dullemond, C. P., \& Paresce, F. 2003, \aap, 400, L21
\bibitem[Wooden et al. (1999)]{wooden99} Wooden, D. H., Harker, D. E., Woodward, C. E., Butner, H. M., Koike, C., Witteborn, F. C., McMurtry, C. W. 1999, \apj, 517, 1034
\bibitem[Weinberger et al. (2003)]{weinberger03}Weinberger, A. J., Becklin, E. E., \& Zuckerman, B. 2003, \apj, 584, L33
\end{thebibliography}
\end{document}